\documentclass[10pt,superscriptaddress,twocolumn,amsmath,nofootinbib,amssymb,aps,prl]{revtex4-1}

\usepackage{graphicx}
\usepackage{dcolumn}
\usepackage{bm}

\usepackage{braket}
\usepackage{color}
\usepackage{xcolor}

\begin{document}

\title{Convergence of the multimode quantum Rabi model of circuit quantum electrodynamics}

\author{Mario F. Gely}
\thanks{These authors contributed equally to this manuscript.}
\affiliation{Kavli Institute of NanoScience, Delft University of Technology,PO Box 5046, 2600 GA, Delft, The Netherlands}

\author{Adrian Parra-Rodriguez}
\thanks{These authors contributed equally to this manuscript.}
\affiliation{Department of Physical Chemistry, University of the Basque Country UPV/EHU, Apartado 644, 48080 Bilbao, Spain}

\author{Daniel Bothner}
 \affiliation{Kavli Institute of NanoScience, Delft University of Technology,PO Box 5046, 2600 GA, Delft, The Netherlands}

\author{Ya.~M.~Blanter}
 \affiliation{Kavli Institute of NanoScience, Delft University of Technology,PO Box 5046, 2600 GA, Delft, The Netherlands}

\author{Sal J. Bosman}
 \affiliation{Kavli Institute of NanoScience, Delft University of Technology,PO Box 5046, 2600 GA, Delft, The Netherlands}

\author{Enrique Solano}

\affiliation{Department of Physical Chemistry, University of the Basque Country UPV/EHU, Apartado 644, 48080 Bilbao, Spain}
\affiliation{IKERBASQUE, Basque Foundation for Science, Maria Diaz de Haro 3, 48013 Bilbao, Spain}

\author{Gary A. Steele}

\affiliation{Kavli Institute of NanoScience, Delft University of Technology,PO Box 5046, 2600 GA, Delft, The Netherlands}

\date{\today}

\begin{abstract}
  Circuit quantum electrodynamics (QED) studies the interaction of artificial atoms, open transmission lines and electromagnetic resonators fabricated from superconducting electronics.
  While the theory of an artificial atom coupled to one mode of a resonator is well studied, considering multiple modes leads to divergences which are not well understood.
  Here, we introduce a first-principles model of a multimode resonator coupled to a Josephson junction atom.
  Studying the model in the absence of any cutoff, in which the coupling rate to mode number $n$ scales as $\sqrt{n}$ for $n$ up to $\infty$, we find that quantities such as the Lamb shift do not diverge due to a natural rescaling of the bare atomic parameters that arises directly from the circuit analysis.  
  Introducing a cutoff in the coupling from a non-zero capacitance of the Josephson junction, we provide a physical interpretation of the decoupling of higher modes in the context of circuit analysis.
  In addition to explaining the convergence of the quantum Rabi model with no cutoff, our work also provides a useful framework for analyzing the ultra-strong coupling regime of multimode circuit QED.
\end{abstract}

\maketitle

\begin{figure}[t]
\centering
\includegraphics[width=0.35\textwidth]{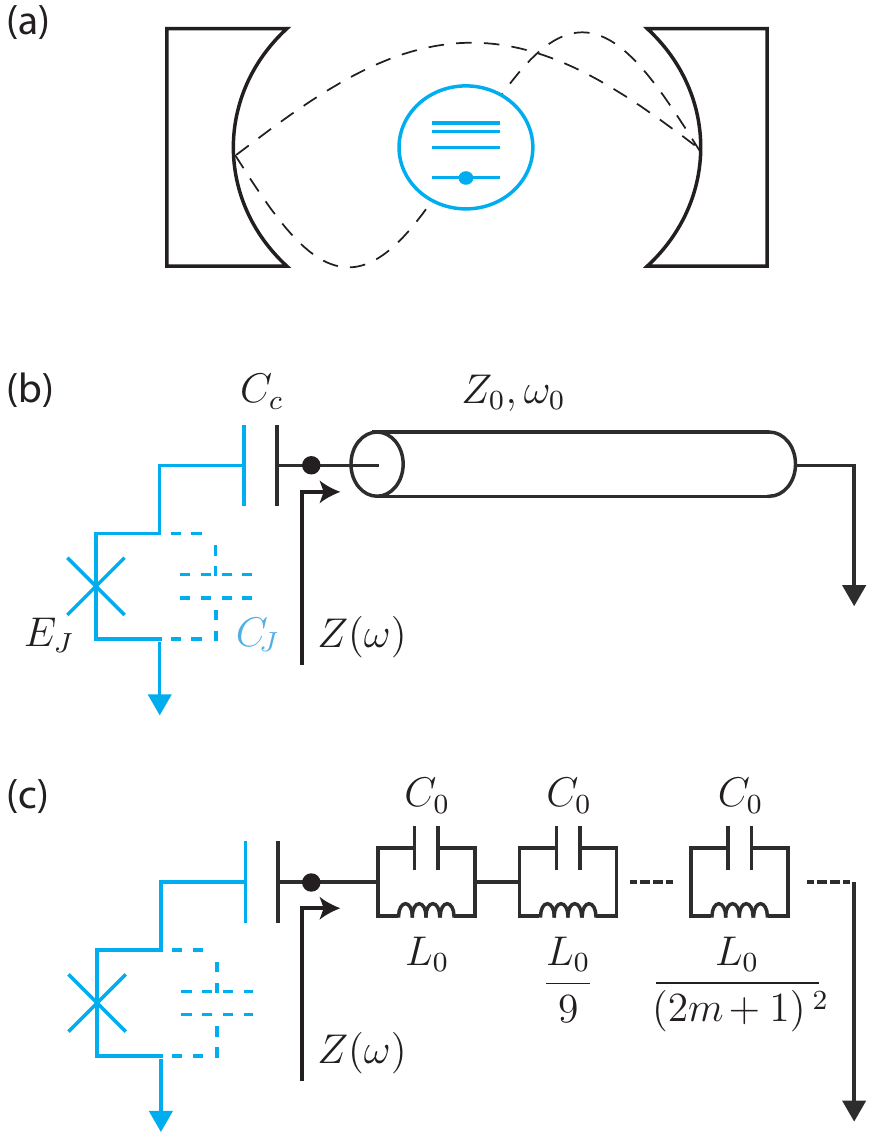}

\caption{(a) Schematic representation of cavity QED: A multilevel atom (in blue) coupled to the electromagnetic cavity modes (dashed black lines). (b) Circuit QED example covered in this work: A Josephson junction anharmonic LC oscillator, or ``artificial atom", coupled to modes of a transmission line. (c) Lumped element equivalent of (b).}
\label{fig:circuit}
\end{figure}

Quantum electrodynamics (QED) explores one of the most fundamental interactions in nature, that of light and matter. In a typical cavity QED scenario, an individual atom interacts through its dipole moment with the electric field of cavity modes, as described by the Rabi model~\cite{rabi_process_1936} and depicted in Fig.~\ref{fig:circuit}(a). One consequence of this interaction is the Lamb shift of the atomic transition frequencies~\cite{Lamb1947,Heinzen1988}. Early attempts at calculating this shift led to the first shortcomings of QED theory, mainly that the transition energies of the atom diverge as the infinite number of electromagnetic modes are considered. Efforts to address these issues gave birth to renormalization theory~\cite{Bethe1947}. Akin to cavity QED is the field of circuit QED~\cite{Wallraff2004}, where artificial atoms such as anharmonic superconducting LC circuits couple to the modes of a waveguide resonator or an open transmission line. Such systems allow the study of a wealth of quantum effects~\cite{Schuster2007,Bishop2009a}, and are one of the most promising platforms for the realization of quantum processors~\cite{Takita2016,Kelly2015a,Riste2015}. Despite experimental successes, ``ad hoc'' multimode extensions of the Rabi model suffer from divergences when considering the limit of infinite modes in a waveguide resonator~\cite{Houck2008,Filipp2011a}.

Aware of this problem, Nigg \textit{et al.}~\cite{Nigg2012,Girvin2011} and Solgun \textit{et al.}~\cite{solgun2014blackbox} developed the method of black-box quantization to obtain effective Hamiltonians with higher predictive power. While a practical tool for weakly anharmonic systems, this method was not designed for systems with strong anharmonicity, such as a Cooper pair box~\cite{Koch2007}. Furthermore, in applying the black-box procedure, the form of the quantum Rabi Hamiltonian is not preserved, and while it gives the correct energy spectrum, it is not clear how to identify and connect it to a coupling rate between the two bipartite systems. Doing so, it is no longer possible to directly identify which parts of the Rabi interaction lead to certain energy shifts of bare quantities, such as the Bloch-Siegert shift, highly relevant for studying the physics of ultra-strongly coupled (USC) systems~\cite{forn-diaz_observation_2010,Exp}. 

In this paper, we derive a first-principles Hamiltonian model addressing these issues. This Hamiltonian is expressed in the basis of the uncoupled resonator modes and the atom, is valid for arbitrary atomic anharmonicities, and allows us to understand why previous attempts at extending the Rabi Hamiltonian have failed. The presence or not of a Josephson capacitance $C_J$ in our study leads to two important results. First, in the limit $C_J\rightarrow 0$, the coupling rates follow a square root increase up to an infinite number of modes. Without introducing a cutoff in the number of coupled modes, we show that a first-principles analysis of the quantum circuit leads to convergence of the energy spectrum. The $C_J=0$ limit also highlights a natural renormalization of Hamiltonian parameters, arising from the circuit analysis, which is essential to understanding how to reach correct multimode extensions of the Rabi model. Secondly, we study the experimentally relevant case $C_J>0$, which introduces a cutoff that suppresses the coupling to higher modes~\cite{DeLiberato2014a,Garcia-Ripoll2015a,malekakhlagh2016origin}. In particular, we provide an analysis of this regime and discuss the physics of this cutoff in the context of a lumped element circuit model. This results in a useful tool for studying multimode circuit QED in the framework of the Rabi Hamiltonian, or for studying strongly anharmonic regimes, out of reach of the black-box quantization method. This model was indispensable in extracting the Bloch-Siegert shift in the experiment of Ref.~\cite{Exp}, where a naive extension of the Rabi model would predict a Lamb shift of more than three times the atomic frequency due to 35 participating modes before any physically-motivated cutoff, such as the qubit's physical size or the Junction capacitance, becomes relevant.

The circuit QED system studied in this work is an artificial atom (AA) formed from an anharmonic LC oscillator~\cite{Koch2007}, capacitively coupled to a quarter-wave ($\lambda/4$) transmission line resonator~\cite{Pozar2009} as depicted in Fig.~\ref{fig:circuit}(b). The AA is a superconducting island connected to ground through a Josephson junction characterized by its Josephson energy $E_J$. It has a capacitance to ground $C_J$ and is coupled to the voltage anti-node of the resonator, with characteristic impedance $Z_0$ and fundamental mode frequency $\omega_0/2\pi$, through a capacitance $C_c$. The Josephson junction acts as a non-linear inductor, providing a source of single-photon anharmonicity in the oscillations of current flowing through it. In order to clearly illustrate the most novel aspect of our model, the renormalization of the charging energy, we first consider the case $C_J=0$. Despite the absence of a cutoff in the number of coupled modes in this case, we find that the energy spectrum still converges. The case of $C_J>0$ is discussed at the end of this article and in detail in the Supplementary Material~\cite{SI}.

\textit{Circuit Hamiltonian} --- We consider a Hamiltonian in which each uncoupled harmonic mode of the resonator, with resonance frequency $\omega_m$ and annihilation operator $\hat{a}_m$, is coupled to the transition between the bare atomic states $\ket{i}$, $\ket{j}$ with energies $\hbar\epsilon_i$, $\hbar\epsilon_j$  through a coupling strength $\hbar g_{m,i,j}$~\cite{Zhu2013,Sundaresan2015}. We derive such a Hamiltonian by constructing a lumped element equivalent circuit, or Foster decomposition, of the transmission line resonator as represented in Fig.~\ref{fig:circuit}(c). The input impedance of a shorted transmission line, at a distance $\lambda/4$ from the short, $Z(\omega) = iZ_0\tan (\pi\omega/(2\omega_0))$, is equal to that of an infinite number of parallel LC resonators with capacitances $C_0 = \pi/(4\omega_0 Z_0)$ and inductances $L_m = 4Z_0/((2m+1)^2\pi \omega_0)$. In order to consider a finite number of modes $M$ in the model, one replaces the $m\ge M$ LC circuits in Fig.~\ref{fig:circuit}(c) by a short circuit to ground. This removes the $m\ge M$ resonances in the resonator input impedance $Z(\omega)$ with little effect on $Z(\omega)$ for $\omega \ll \omega_M$. The focus of this paper is on the evolution of the Hamiltonian parameters as a function of this system size $M$, and the consequences on the energy spectrum. Using the tools of circuit quantization~\cite{Devoret1997}, we obtain as the Hamiltonian of the system~\cite{SI}

\begin{figure*}[t]
\includegraphics[width=0.8\textwidth]{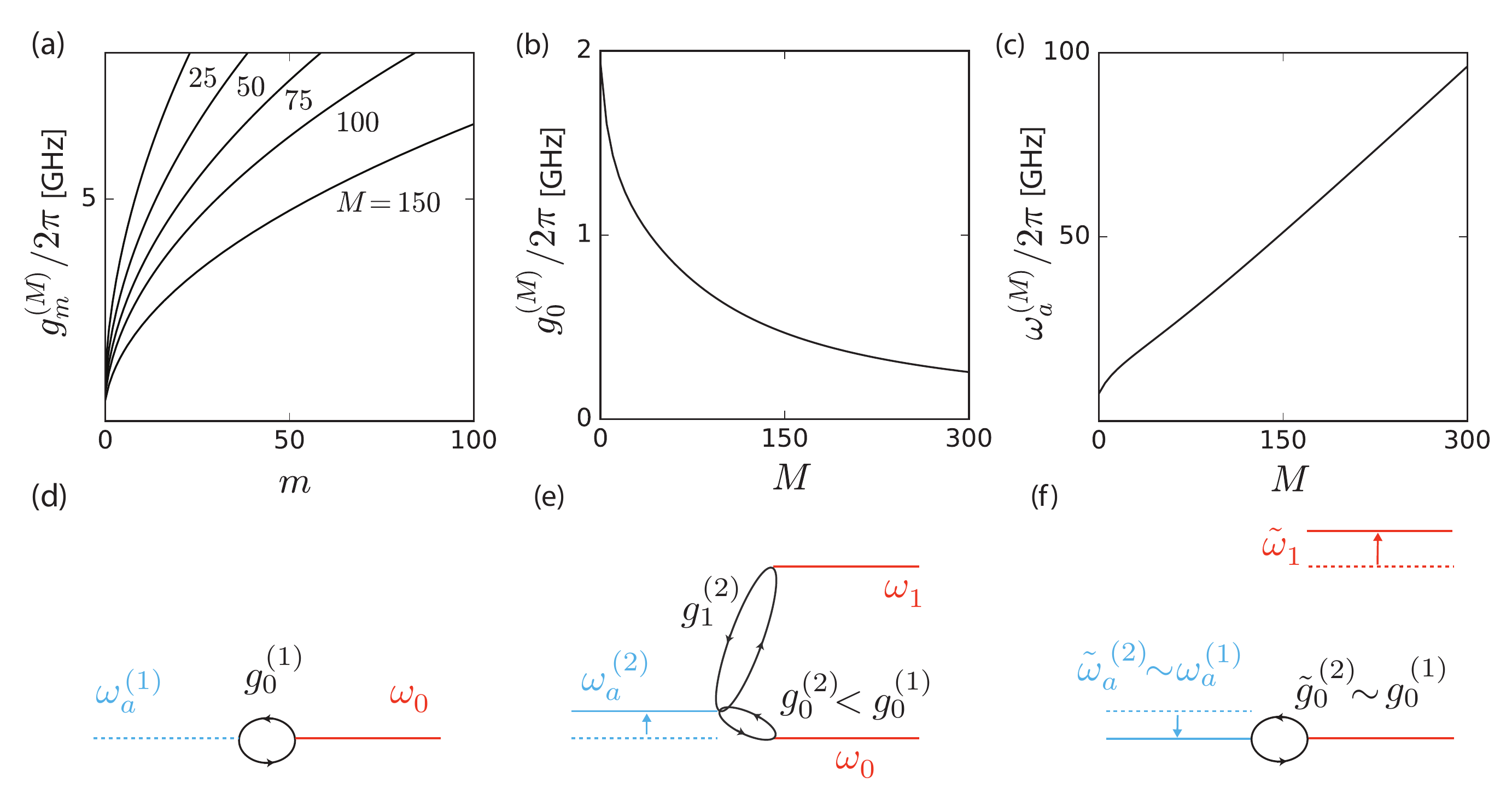}
\caption{Renormalization of the Hamiltonian parameters and dressing of the atom by higher modes. For the plots, we choose $\omega_0/2\pi = 10$ GHz, $Z_0=50\ \Omega$, $C_c=50$ fF and $E_J/h=20$ GHz. Although this places the AA in the transmon limit, we note that the scaling shown in the figure is exact for all regimes, including the Cooper pair box limit. (a) Coupling of the ground to first excited state transition of the atom $\ket{g}\rightarrow\ket{e}$ to the resonator modes as a function of the mode number $m$: $\hbar g_{m}^{(M)} = \sqrt{2m+1}V_{\text{zpf,0}}2e\bra{g}^{(M)}\hat{N}_J\ket{e}^{(M)}$ for different values of $M$. The $M$-dependence of the coupling is detailed in (b). (b),(c) Renormalization of the coupling strength and the atomic frequency through the change in charging energy $E_C^{(M)}$ as a function of $M$. For large $M$, the coupling diminishes with $1/M$ and the atomic frequency increases linearly with $M$. (d-f) Schematic energy diagrams of the renormalization procedure in the case of the atom and fundamental mode at resonance. (a) $M=1$ (b) $M=2$.  Adding a mode shifts the bare atomic energy upwards and changes the values of the couplings. (f) 
  Dressing the atom with the second mode results in a dressed atomic state with atomic frequency and coupling to the $m=1$ mode close to that of the $M=1$ model.}
\label{fig:renormalization}
\end{figure*}

\begin{align}
\begin{split}
  \hat{H}^{(M)} &= \sum_i\hbar\epsilon_i^{(M)} \ket{i}^{(M)}\bra{i}^{(M)}  + \sum_{m=0}^{m<M}\hbar\omega_m\hat{a}_m^\dagger \hat{a}_m \\
  &+ \sum_{i,j}\sum_{m=0}^{m<M}\hbar g^{(M)}_{m,i,j}\ket{i}^{(M)}\bra{j}^{(M)}(\hat{a}_m+\hat{a}_m^\dagger)\ .
  \end{split}
  \label{eq:circuit_hamiltonian}
\end{align}
The eigenfrequencies of the higher resonator modes $\omega_m$ are related to that of the fundamental mode through $\omega_m = (2m+1)\omega_0$. The coupling strength $\hbar g^{(M)}_{m,i,j} = 2e V_{\text{zpf},m}\bra{i}^{(M)}\hat{N}_J\ket{j}^{(M)}$ scales with the square root of the mode number $m$ through the zero-point voltage fluctuations of the $m$-th mode $V_{\text{zpf},m}=\sqrt{2m+1}\sqrt{\hbar\omega_0/2C_0}$.
Since we will concentrate on the frequency and coupling of the first atomic transition $\ket{g}\rightarrow\ket{e}$, we use the shorthand $\omega_a^{(M)} = \epsilon_e-\epsilon_g$ and $ g_{m}^{(M)} = g_{m,g,e}^{(M)}$ throughout this paper.

The (bare) AA eigenstates $\ket{i}^{(M)}$ and energies $\hbar \epsilon_i^{(M)}$ in Eq.\ (\ref{eq:circuit_hamiltonian}) are those that diagonalize the Hamiltonian
\begin{equation}
    \hat{H}^{(M)}_{\text{AA}} = 4E_C^{(M)} \hat{N}_J -E_J \cos(\hat{\delta})\ .
    \label{eq:CPB_hamiltonian}
\end{equation}
Here $\hat{N}_J$ is the quantum number of Cooper-pairs on the island conjugate to $\hat{\delta}$ the superconducting phase difference across the junction, and $E_C^{(M)}$ is the charging energy of the island. This choice of the decomposition of the Hamiltonian is one in which the bare atom corresponds to purely anharmonic degrees of freedom (currents flowing only through the junction) and the bare cavity to purely harmonic degrees of freedom (currents flowing only through the linear cavity inductors). 

The crucial consequence of quantizing our model is a renormalization of the parameters of the Hamiltonian as modes are added. In particular, the charging energy $E^{(M)}_C$ of the (bare) AA depends explicitly on the number of modes included in the equivalent circuit
\begin{equation}
  E^{(M)}_{C} = \frac{e^2}{2}\frac{C_0+MC_c}{C_0C_c}\ ,
  \label{eq:Ec}
\end{equation}
as reported previously in the case of multiple atoms coupled to a single mode~\cite{jaako2016ultrastrong}, or in the context of metallic dots coupled to a resonator~\cite{bergenfeldt2012microwave}. For the case of $M \rightarrow \infty$ with $C_J = 0 $, the charging energy of the bare atom diverges. This divergence arises from the definition of the bare atom as current oscillations flowing only through the junction. As $M \rightarrow \infty$, the impedance path through only the series capacitors of the resonator equivalent circuit diverges. Charge from currents through the junction can no longer oscillate on $C_c$ and $\omega_a^{(M)}$ diverges. For the case of $M=1$ and $C_c\ll C_0$, Eq.~(\ref{eq:Ec}) simplifies to the standard definition of the charging energy $E_C = e^2/2C_c$~\cite{Koch2007}. With $M > 1$, we will see that a more complex picture emerges.

\textit{Renormalization of the atomic parameters} --- In Fig.~\ref{fig:renormalization}, we explore the renormalization of the parameters of our model as the number of modes $M$ is increased. Through the change in charging energy, both the eigenstates $\ket{i}^{(M)}$ and coupling strengths $g_m^{(M)}$ depend on $M$. For a fixed number of modes $M$, the coupling $g_m^{(M)}$ of the atom to mode $m$ scales with the square root of the mode number $m$: $g_m^{(M)} = g_0^{(M)}\sqrt{2m+1}$. From this coupling, each mode will induce a Lamb shift of the atomic energy $\chi_m\simeq -2 \big(g_m^{(M)}\big)^2/\omega_m = -2 \big(g_0^{(M)}\big)^2/\omega_0$, a formula valid in the transmon regime when $\omega_m$ is much larger than the bare atomic frequency. With the typical assumption of coupling and bare atomic frequency independent of $M$, summing the Lamb shifts of every mode would lead to diverging values of the dressed atomic frequencies. This leads to the divergences found in typical multimode extensions of the quantum Rabi model.

In the model presented here, however, we find that the full quantization of the lumped-element circuit leads to a Hamiltonian in which both the bare atomic frequency $\omega_a$ and the couplings to the modes $g_m$ are explicitly dependent on the number of modes $M$ included in the model. As the number of modes $M$ in the model increases, the bare atomic couplings $g_m^{(M)}$ are suppressed (Fig.~\ref{fig:renormalization}(a,b)), and the bare atomic frequency $\omega_a^{(M)}$ increases (Fig.~\ref{fig:renormalization}(c)), diverging for an infinite number of modes. As we will see, however, convergence is obtained in the dressed transition energy of the atom when including the Lamb-shift from higher modes of the resonator.

As an illustration of how renormalization in our model leads to convergence of the spectrum, let us consider the case shown in Fig.~\ref{fig:renormalization}(d--e) in which the fundamental mode is resonant with the atomic frequency $\omega_a^{(1)}$ when $M=1$ (Fig.~\ref{fig:renormalization}(d)).  Including an additional mode with frequency $\omega_1$ will lead to an upwards shift of the bare atomic transition $\omega_{a}^{(1)}\rightarrow\omega_{a}^{(2)} > \omega_{a}^{(1)}$ and a change of the coupling $g^{(1)}_{0}\rightarrow g^{(2)}_{0}<g^{(1)}_{0}$ through the renormalization of the charging energy (Fig.~\ref{fig:renormalization}(e)). Diagonalizing the subsystem of the atom and mode 1 in our model, the transition energy of the atom is shifted down again near resonance with the fundamental mode $\omega_{a}^{(2)}\rightarrow \tilde{\omega}_{a}^{(2)}\sim \omega_{a}^{(1)}$ by the dispersive shift, and the coupling of the atomic transition to the fundamental mode is increased $g_{0}^{(2)}\rightarrow\tilde{g}_{0}^{(2)}\sim g^{(1)}_{0}$ (Fig.~\ref{fig:renormalization}(f)). In this way, the resulting vacuum Rabi splitting of the fundamental mode is found to be similar to that of the $M=1$ model, despite the decrease in the bare coupling rates $g_0^{(2)}$.

Note that in our model, the value of $E_J/E_C^{(M)}$ of the bare atom, which determines its anharmonicity~\cite{Koch2007}, is also a function of $M$. It would seem that in the limit $M \rightarrow \infty$, the bare atom would be deep in the Cooper pair box limit. However, including the hybridization with the cavity, the low energy sector of $\hat{H}^{(M+1)}$ is well approximated by a model with $M$ modes where the charging energy is not $E_C^{(M)}$ but
\begin{equation}
\tilde{E}_C^{(M)} = E_C^{(M+1)} - \hbar(\bar{g}^{(M+1)}_{M})^{2}/4\omega_{M}\ ,
\label{eq_EC_tilde_M}
\end{equation}
where $\bar{g}^{(M)}_m$ is the coupling constant without the dipole moment $\bra{i}^M\hat{N}_J\ket{j}^M$. For this to hold, $\hbar \omega_M$ must be larger than the characteristic energy of the low energy sector of $\hat{H}^{(M)}$. In this case, the vacuum of the $(M+1)$-th mode, shifted by $\left(\bar{g}^{(M+1)}_{M}/\omega_{M}\right)\hat{N}_J$, is a good variational choice for the low lying energy sector of $\hat{H}^{(M+1)}$. In this subspace, the effective Hamiltonian is of the same form as $\hat{H}^{(M)}$, but with charging energy $\tilde{E}_C^{(M)}$~\cite{SI}. This result matches with the zero-th order of a Schrieffer-Wolff approximation~\cite{Schrieffer1966,Bravyi2011}. We can iterate this procedure to a mode $L$. For $M\rightarrow\infty$, an effective Hamiltonian with $L$ modes will have a finite charging energy 
\begin{equation}
\tilde{E}_C^{(L)} = \lim_{M\rightarrow\infty}E_C^{(M)} - \sum_{m\geq L}^{M}\hbar(\bar{g}^{(M)}_{m})^{2}/4\omega_{m} .\label{eq_EC_tilde_LM}
\end{equation}
The interaction with higher modes therefore modifies the charging energy of the dressed atom, leading to a convergence of the atomic anharmonicity as well. This formula applies for all values of $C_J$, but for $C_J=0$, we have $\tilde{E}_C^{(M)} = E_C^{(M)}$, \textit{i.e.}, the dressing from higher modes exactly compensates the renormalization of the charging energy.

\begin{figure}[t]
\includegraphics[width=0.5\textwidth]{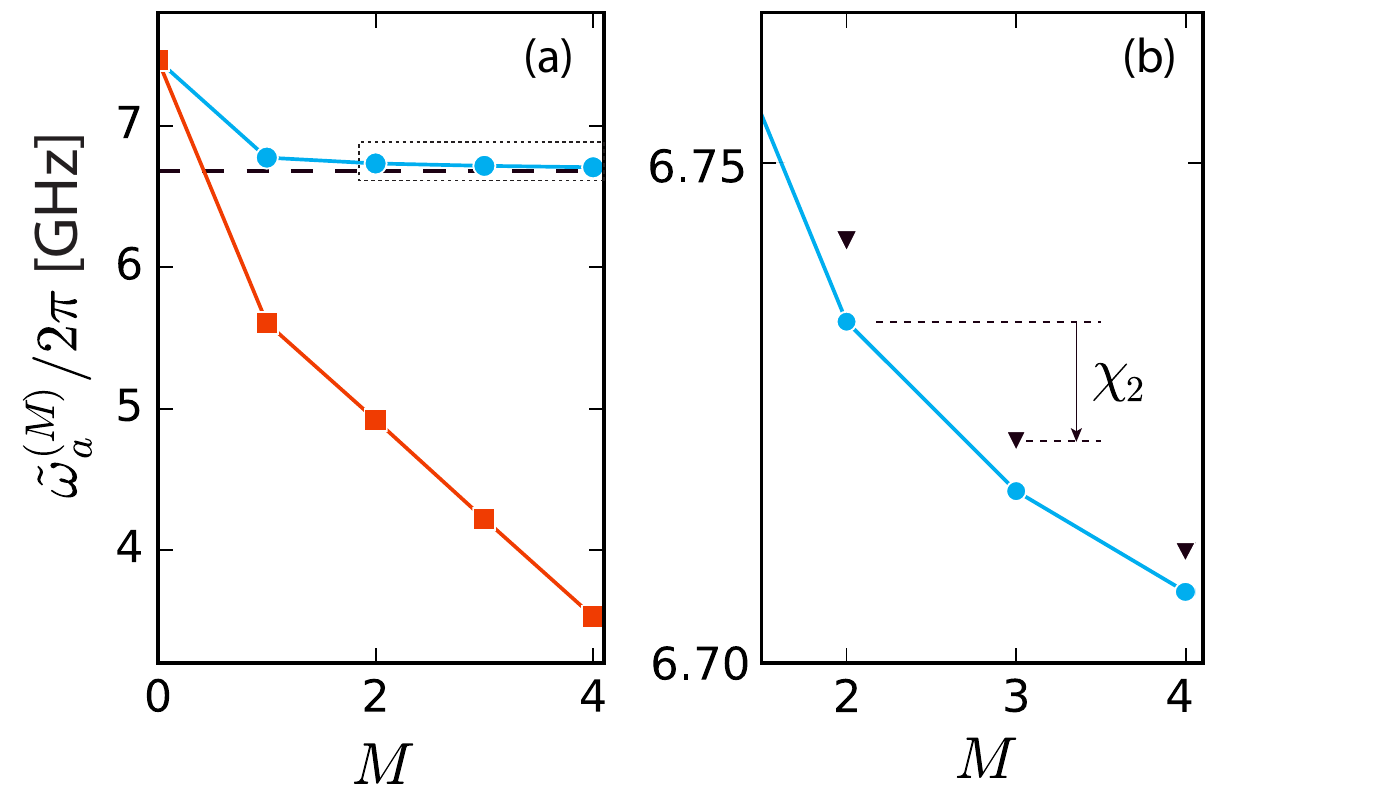}
\caption{Calculated spectrum as a function of the number of modes $M$ included in the model.  (a) Dots (squares) correspond to a diagonalization of the circuit (non-)renormalized extended-Rabi model. The frequency obtained by black-box quantization of the circuit Fig.~\ref{fig:circuit}(b) (dashed line) provides a point of reference corresponding to the case when all modes are included. (b) Zoom of dashed box in (a). Triangles show the prediction based on a first order classical approximation of the Lamb shift: $\chi_M = \tilde{\omega}_{a}^{(M+1)}-\tilde{\omega}_{a}^{(M)} \simeq -2 ((g_{M}^{(0)})^2/\omega_M)((\omega_{a}^{(0)})^{2}/\omega_M^2)$~\cite{SI}.}
\label{fig:lambshift}
\end{figure}

In order to illustrate the effectiveness of this renormalization, in Fig.~\ref{fig:lambshift} we compare a diagonalization of Hamiltonian (\ref{eq:circuit_hamiltonian}) to a non-renormalized multimode extension of the quantum Rabi model, implemented by removing the $M$-dependence of the charging energy $E_C^{(M)}\rightarrow e^2/(2C_c)$. The dashed line indicates the result of the black-box quantization (BB) method~\cite{Nigg2012} as a point of reference. The calculations are performed using the same physical parameters as in Fig.~\ref{fig:renormalization}. Compared to the non-renormalized model, which diverges linearly, a diagonalization of the first-principle Hamiltonian (\ref{eq:circuit_hamiltonian}) converges towards the value expected from BB.

It is also interesting to note that the corrections from our model are non-perturbative: perturbation theory fails to give a value for the Lamb shift resulting from including an extra mode. Using a circuit analysis of coupled LC oscillators (see~\cite{SI}), in the transmon regime, $E_J \gg E_C^{(M)}$, we find an estimate of the shift in the dressed AA energy when including an additional mode in the model given by $\chi_m\simeq -2(g_m^2/\omega_m)(\tilde{\omega}_{a}^2/\omega_m^2)$. This formula can be used to estimate the number of relevant modes to include in a simulation, and can be though of as a replacement of the usual expression for the Lamb shift $\chi_m^{\text{Lamb}}\simeq -2 g_m^2/\omega_m$.

\begin{figure}[t]
\centering
\includegraphics[width=0.4\textwidth]{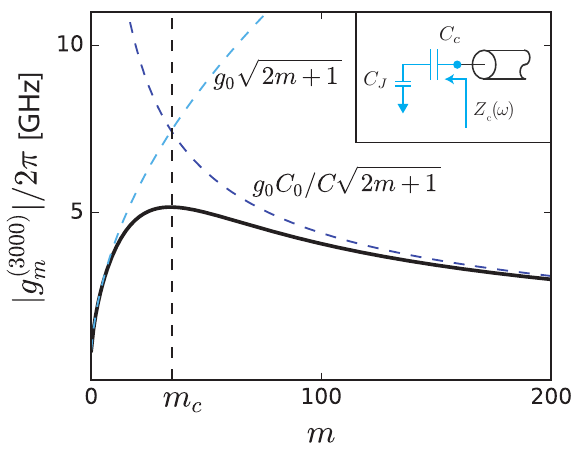}
\caption{High-frequency cutoff for $C_J \neq 0$. The capacitive loading at the left boundary of the resonator shown in the inset transforms this point from a voltage anti-node to a voltage node for higher modes. The mode $m_c\simeq 35$ marks this transition. The solid line corresponds to the coupling strength as a function of the number of modes. With $C_J \neq 0$ the coupling strength converges to a non-zero value for large $M$ hence the choice $M=3000$. Dashed lines: asymptotic values of the coupling, with $g_0 = g_0^{(M=3000)}$ and $C = C_cC_J/(C_c+C_J)$.}
\label{fig:cutoff}
\end{figure}

\textit{Consequence of introducing a high-frequency cutoff} --- In a realistic system, higher modes will tend to decouple from the atom due to several coexisting physical mechanisms~\cite{Filipp2011a}. One such mechanism is the capacitance of the Josephson junction $C_J$. In particular, the capacitive loading of the cavity from the AA illustrated in the inset of Fig.~\ref{fig:cutoff} leads to a decreasing impedance to ground $Z_c(\omega) \simeq i(C_J+C_c)/\omega C_J C_c$ at the end of the resonator when $\omega\gg\omega_{a}$. When the mode frequencies become such that this impedance is lower that the characteristic impedance of the resonator $|Z_c(\omega)| \ll |Z_0|$, this voltage anti-node of the resonator, to which the AA couples, becomes a voltage node, and the coupling vanishes. Additionally, the eigen frequencies will span from those of a $\lambda/4$ resonator for the lower modes to those of a $\lambda/2$ resonator $\omega_m \rightarrow 2m\omega_0$ for the higher lying modes.

This effect can be captured with the same quantization procedure applied to the circuit in Fig.~\ref{fig:circuit}(c) with $C_J \ne 0$ and is detailed in the supplementary material~\cite{SI}. Mathematically, the cutoff in the coupling is due to a mode-mode coupling term of the form $\sum_{m=0}^{m<M}\sum_{m'=m+1}^{m<M}G^{(M)}_{m,m'}(\hat{a}_m+\hat{a}_m^\dagger)(a_{m'}+a_{m'}^\dagger)$, which arises naturally from the circuit quantization. This is the equivalent of the $A^2$ term discussed in Refs.~\cite{DeLiberato2014a,Garcia-Ripoll2015a,malekakhlagh2016origin}. Diagonalizing the Hamiltonian of coupled resonator modes leads to decreasing zero-point voltage fluctuations of the modes at the coupling node. As shown in Fig.~\ref{fig:cutoff}, with a capacitance to ground $C_J = 5$ fF close to the experimental parameters of Ref.~\cite{Exp}, the expected cutoff occurs when $|Z_c(\omega_{m_c})| \simeq |Z_0|$, or equivalently at the mode number $m_c \simeq (C_J+C_c)/2\omega_0 Z_0C_JC_c$. This mechanism is accompanied by the appearance of an upper bound in the renormalized charging energy, such that Eq.~(\ref{eq:Ec}) becomes
\begin{equation}
  E^{(M)}_{C} = \frac{e^2}{2}\frac{C_0+MC_c}{MC_cC_J+C_0(C_c+C_J)},
\end{equation}
and $E^{(M)}_{C}\rightarrow e^2/2C_J$ for $M\rightarrow \infty$. We emphasize however, that this cutoff is not a necessary condition for the convergence of the energy spectrum: the model described above with $C_\textrm{J} = 0$ converges even in the absence of such a cutoff. This is to be contrasted with typical models of (natural) atoms coupled to cavity modes where high frequency cutoffs must be imposed to obtain finite predictions~\cite{Mandel1995}. It would be interesting to study if the ideas developed in this work apply to such systems. 

\textit{Conclusion} --- 
We have developed a first-principles multimode quantum Rabi model of circuit QED from a compact lumped element equivalent circuit.
Using this formulation, we derived the convergence of quantities such as the Lamb shift in the absence of any high frequency cutoff, arising from a natural renormalization of the Hamiltonian parameters as modes are added. 
We also study the implications of a finite junction capacitance, which introduces a cutoff in the coupling to high frequency modes, but does not change the renormalization that occurs when additional modes are included in the circuit.
For both cases with and without a junction capacitance, we show that when constructing a quantum Rabi model from this compact lumped element equivalent circuit, it is crucial to include this renormalization to get correct Hamiltonian parameters from the values of the circuit elements.
This work provides a useful framework for an intuitive understanding and modeling of experiments in the multimode ultra-strong coupling regime.
This formulation of the multimode quantum Rabi model in the context of circuits hints at an intuitive picture on how this renormalization can arise physically, and it suggests the study of how this proposed physical picture could be applied to other problems in quantum field theory.

\textit{Acknowledgments} A. P-R and E. S. thank Enrique Rico and \'I\~nigo Egusquiza for useful discussions. M. G. and G. S. thank  Yuli V. Nazarov for useful discussions. The authors acknowledge funding from UPV/EHU UFI 11/55, Spanish MINECO/FEDER FIS2015-69983-P, Basque Government IT986-16 and PhD grant PRE-2016-1-0284, the Netherlands Organization for Scientific Research (NWO), the Dutch Foundation for Fundamental Research on Matter (FOM), and the European Research Council (ERC).

\textit{Note} After we finished this manuscript, we became aware of Ref.~\cite{malekakhlagh2017cutoff}, which arrives at a similar conclusions as the last section of this paper through a different approach.

\bibliography{library,extra}

\end{document}



\title{Supplementary Material: Divergence-free multi-mode circuit quantum electrodynamics}

\author{Mario F. Gely}
\thanks{These authors contributed equally to this manuscript.}
\affiliation{Kavli Institute of NanoScience, Delft University of Technology,PO Box 5046, 2600 GA, Delft, The Netherlands}

\author{Adrian Parra-Rodriguez}
\thanks{These authors contributed equally to this manuscript.}
\affiliation{Department of Physical Chemistry, University of the Basque Country UPV/EHU, Apartado 644, 48080 Bilbao, Spain}

\author{Daniel Bothner}
 \affiliation{Kavli Institute of NanoScience, Delft University of Technology,PO Box 5046, 2600 GA, Delft, The Netherlands}

\author{Ya.~M.~Blanter}
 \affiliation{Kavli Institute of NanoScience, Delft University of Technology,PO Box 5046, 2600 GA, Delft, The Netherlands}

\author{Sal J. Bosman}
 \affiliation{Kavli Institute of NanoScience, Delft University of Technology,PO Box 5046, 2600 GA, Delft, The Netherlands}

\author{Enrique Solano}

\affiliation{Department of Physical Chemistry, University of the Basque Country UPV/EHU, Apartado 644, 48080 Bilbao, Spain}
\affiliation{IKERBASQUE, Basque Foundation for Science, Maria Diaz de Haro 3, 48013 Bilbao, Spain}

\author{Gary A. Steele}

\affiliation{Kavli Institute of NanoScience, Delft University of Technology,PO Box 5046, 2600 GA, Delft, The Netherlands}

\date{\today}

\maketitle
\onecolumngrid
\title{Supplementary material}

\maketitle

\section{Derivation of the circuit Hamiltonian}

The input impedance of a shorted transmission line, at a distance $\lambda_0/4$ from the short (see Ref.~\cite{Pozar2009}) is given by
\begin{equation}
  Z(\omega) = i Z_0 \tan\bigg(\frac{\pi}{2}\frac{\omega}{\omega_0}\bigg)\ ,
\end{equation}
where $Z_0$ is the characteristic impedance of the waveguide, $\omega_0/2\pi$ is the resonance frequency and $\lambda_0$ the wavelength of the fundamental mode of the quarter wave resonator when the AA is replaced by an open termination. The partial fraction expansion of the tangent
\begin{equation}
  \tan(z) = \sum_{m=0}^\infty \frac{-2z}{z^2-(m+\frac{1}{2})^2\pi^2}
\end{equation}
leads to an expression for the resonators imput impedance which is equal to that of an infinite number of parallel LC resonators. Each of them corresponds to a resonance mode
\begin{equation}
  Z(\omega) = i Z_0 \tan\bigg(\frac{\pi}{2}\frac{\omega}{\omega_0}\bigg) = \sum_{m=0}^\infty \frac{1}{iC_0\omega+\frac{1}{iL_m\omega}}\ ,
\end{equation}
\begin{equation}
  C_0 = \frac{\pi}{4\omega_0 Z_0}\ ,
\end{equation}
\begin{equation}
  L_m = \frac{1}{(2m+1)^2}\frac{4Z_0}{\pi \omega_0}\ .
\end{equation}

\begin{figure}[b]
\centering
\includegraphics[width=0.5\textwidth]{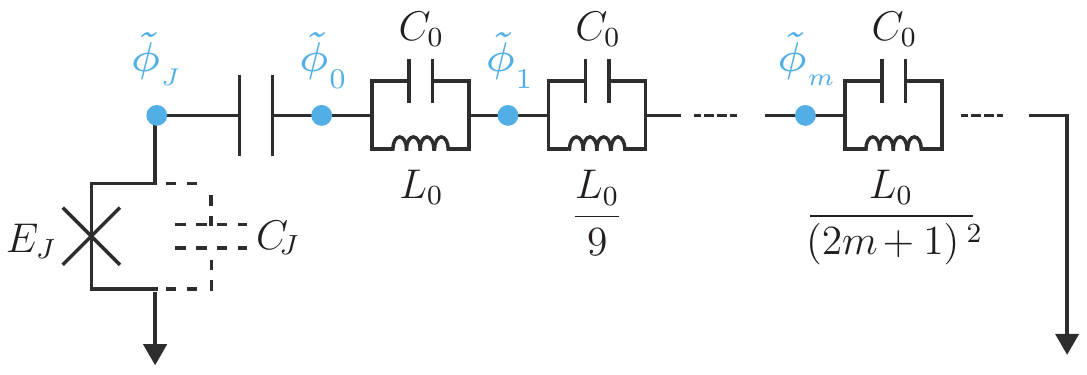}
\caption{Circuit quantized in this section. In accordance with Ref.~\cite{Devoret1997}, the degrees of freedom of the circuit are chosen to be the fluxes (indicated in blue) at the nodes of the circuit.}
\label{fig:circuit_notations}
\end{figure}

We truncate the system to the first $M$ resonators and use the tools of circuit quantization to obtain the corresponding Hamiltonian. Following the methodology given in Refs.~\cite{Devoret1997} and \cite{Girvin2011}, we start by defining a set of nodes of the circuit and their corresponding fluxes. We define the flux $\tilde{\phi}$ from the voltage $v$ of that node to ground as
\begin{equation}
  \tilde{\phi}(t) = \int_{-\infty}^t v(t')dt'\ .
\end{equation}
 As described in Fig. \ref{fig:circuit_notations}, the node corresponding to the superconducting island of the AA is denoted by the subscript $J$, and we number from $0$ to $M-1$ the nodes corresponding to the fluxes from the $m$-th LC oscillator to the coupling capacitor. The Lagrangian of the system is given by
 \begin{equation}
 \begin{split}
   \mathcal{L} &= C_J\frac{\dot{\tilde{\phi}}_J^2}{2} + C_c\frac{(\dot{\tilde{\phi}}_J-\dot{\tilde{\phi}}_0)^2}{2} + \sum_{m=0}^{m<M-1} C_0\frac{(\dot{\tilde{\phi}}_m-\dot{\tilde{\phi}}_{m+1})^2}{2} + C_0\frac{(\dot{\tilde{\phi}}_{M-1})^2}{2} \\
   &+E_J\cos\bigg(2\pi\frac{\tilde{\phi}_J}{\Phi_0}\bigg) - \sum_{m=0}^{m<M-1} (2m+1)^2\frac{(\tilde{\phi}_m-\tilde{\phi}_{m+1})^2}{2L_0} - (2M-1)^2\frac{(\tilde{\phi}_{M-1})^2}{2L_0}\ ,
   \end{split}
 \end{equation}
where $\Phi_0 = h/2e$ corresponds to the flux quantum and is not to be confused with $\tilde{\phi}_0$. We now make the change of variables $\phi_m = \tilde{\phi}_{m} - \tilde{\phi}_{m+1}$ for $0\le m<M-1$, leaving the remaining two variables unchanged $\phi_{M-1} = \tilde{\phi}_{M-1}$ and $\phi_J = \tilde{\phi}_J$. The Lagrangian then reads
 \begin{equation}
 \begin{split}
   \mathcal{L} &= C_J\frac{\dot{\phi}_J^2}{2} + C_c\frac{\big(\dot{\phi}_J-\sum_{m=0}^{m<M}\dot{\phi}_m\big)^2}{2} + \sum_{m=0}^{m<M} C_0\frac{(\dot{\phi}_m)^2}{2}\\
   &+E_J\cos\bigg(2\pi\frac{\phi_J}{\Phi_0}\bigg) - \sum_{m=0}^{m<M} (2m+1)^2\frac{(\phi_m)^2}{2L_0}\ .
   \end{split}
 \end{equation}
Now the variables $\dot{\phi}_m$ correspond directly to the voltage difference across the capacitance of the $m$-th LC oscillator. With the objective of writing a Hamiltonian, it is useful to express the capacitive part of the Lagrangian in matrix notation
\begin{equation}
  \mathcal{L} = \frac{1}{2}\boldsymbol{\dot{\phi}}^T \boldsymbol{C} \boldsymbol{\dot{\phi}} +E_J\cos\bigg(2\pi\frac{\phi_J}{\Phi_0}\bigg) - \sum_{m=0}^{m<M} (2m+1)^2\frac{(\phi_m)^2}{2L_0}\ ,
\end{equation}
\begin{equation}
  \boldsymbol{\dot{\phi}}^T = 
  \begin{bmatrix}
 \dot{\phi}_J    & \dot{\phi}_0     &  \dot{\phi}_1  &  \dot{\phi}_2    & \cdots & \dot{\phi}_{M-1}
\end{bmatrix}\ ,
\end{equation}

\begin{equation}
  \boldsymbol{C} = 
\begin{bmatrix}
 C_J+C_c     & -C_c    &  -C_c     & -C_c      & \cdots & -C_c \\
 -C_c      &  C_0+C_c     & C_c & C_c &  \cdots & C_c  \\
-C_c      &  C_c     & C_0+C_c & C_c &  & \\
 -C_c      &  C_c     & C_c & C_0+C_c &  & \\
 \vdots & \vdots &  &        &     \ddots &\\
 -C_c      &  C_c     &  &  &  & C_0+C_c\\
\end{bmatrix}\ .
\end{equation}
The canonical momenta (dimensionally charges) are equal to
\begin{equation}
  q_i = \frac{\partial \mathcal{L} }{\partial \dot{\phi}_i} = C_{ij}\dot{\phi}_i
\end{equation}
using Einstein summation convention for repeated indices. The Hamiltonian $H = q_i\dot{\phi}_i - \mathcal{L}$ is then given by
\begin{equation}
  H = \frac{1}{2}\boldsymbol{q}^T\boldsymbol{C}^{-1}\boldsymbol{q}  -E_J\cos(2\pi\frac{\phi_J}{\Phi_0}) + \sum_{m=0}^{m<M} (2m+1)^2\frac{(\phi_m)^2}{2L_0}\ ,
\end{equation}
\begin{equation}
  \boldsymbol{q}^T = 
  \begin{bmatrix}
 q_B    & q_0     &  q_1  &  q_2    & \cdots & q_{M-1}
\end{bmatrix}
\end{equation}
and the inverse of the capacitance matrix is
\begin{equation}
\begin{split}
  \boldsymbol{C}^{-1} &= \frac{1}{C_0(MC_cC_J+C_0(C_c+C_J))}\\
  &\times
\begin{bmatrix}
 C_0^2+MC_0C_c    & C_0C_c     &  C_0C_c       & \cdots  \\
C_0C_c    &  C_0(C_c+C_J)+(M-1)C_cC_J    & -C_JC_c  &   \cdots   \\
C_0C_c    &  -C_JC_c     & C_0(C_c+C_J)+(M-1)C_cC_J &   \\
 \vdots & \vdots &  &           \ddots \\
\end{bmatrix}\ .
\end{split}
\end{equation}
It is easy to check this result in a very general way by veryfing that $\boldsymbol{CC}^{-1} = \boldsymbol{C}^{-1}\boldsymbol{C} = \boldsymbol{I}$. We now quantize the canonical variables $q_i\rightarrow\hat{q_i}$, $\phi_i\rightarrow\hat{\phi_i}$, postulating the commutation relation

\begin{equation}
  [\hat{\phi}_i,\hat{q}_j] = i\hbar\delta_{ij}\ .
\end{equation}
This results in the Hamiltonian
\begin{equation}
  \hat{H}^{(M)} = \hat{H}^{(M)}_{\text{AA}} + \hat{H}^{(M)}_{\text{cav}} + \hat{H}^{(M)}_{\text{int}}\ .
\end{equation}
The AA Hamiltonian is defined as
\begin{equation}
  \hat{H}^{(M)}_{\text{AA}} = \frac{1}{2C^{(M)}_{\text{AA}}}\hat{q}_B^2 -E_J\cos\bigg(2\pi\frac{\phi_J}{\Phi_0}\bigg)\ ,
\end{equation}
where the atoms capacitance is given by
\begin{equation}
  C^{(M)}_{\text{AA}} = \frac{C_0(MC_cC_J+C_0(C_c+C_J))}{C_0^2+MC_0C_c}\ .
\end{equation}
Usually, the charge is expressed in number of Cooper pairs $\hat{q}_B = 2e\hat{N}_J$ and the charging energy is given by $E_C^{(M)} = e^2/2C^{(M)}_{\text{AA}}$, resulting in the Hamiltonian
  \begin{equation}
    \hat{H}^{(M)}_{\text{AA}} = 4E_C^{(M)}\hat{N}_J^2 - E_J\cos(2\pi\frac{\hat{\phi}_J}{\Phi_0})\ .
  \end{equation}
In the main text, we introduced the superconducting phase difference accross the junction as $\hat{\delta} = 2\pi\frac{\hat{\phi}_J}{\Phi_0}$.\\
The cavity Hamiltonian is equal to
\begin{equation}
   \hat{H}^{(M)}_{\text{cav}} = \sum_{m =0}^{m<M}\frac{1}{2C^{(M)}_0}\hat{q}_m^2  + \sum_{m=0}^{m<M} (2m+1)^2\frac{(\phi_m)^2}{2L_0}\ ,
\end{equation}
where the effective capacitance of each oscillator is given by
\begin{equation}
  C_0^{(M)} = \frac{C_0(MC_cC_J+C_0(C_c+C_J)}{(M-1)C_cC_J+C_0(C_c+C_J)}\ .
\end{equation}
We define the creation and annihilation operators
\begin{equation}
  \hat{\phi}_m = \frac{-i}{\sqrt{2m+1}}\sqrt{\frac{\hbar}{2}\sqrt{\frac{L_0}{C_0^{(M)}}}}(\hat{a}_m-\hat{a}_m^\dagger)\ ,
\end{equation}
\begin{equation}
  \hat{q}_m = \sqrt{2m+1}\sqrt{\frac{\hbar}{2}\sqrt\frac{C_0^{(M)}}{L_0}}(\hat{a}_m+\hat{a}_m^\dagger)\ ,
\end{equation}
reducing the cavity Hamiltonian to 
\begin{equation}
   \hat{H}^{(M)}_{\text{cav}} = \sum_{m =0}^{m<M}\hbar\omega_m^{(M)}\hat{a}^\dagger\hat{a}\ ,
\end{equation}
\begin{equation}
   \omega_m^{(M)} = \frac{2m+1}{\sqrt{L_0C_0^{(M)}}}\ ,
\end{equation}
where we have dropped the constant energy contributions $\hbar\omega_m^{(M)}/2$. The quantum voltage of each mode is
\begin{equation}
  \hat{V}_m^{(M)} = \frac{\hat{q}_m}{C_0^{(M)}} = V_{\text{zpf},m}^{(M)}(\hat{a}_m+\hat{a}_m^\dagger) ,
\end{equation}
defining the zero point fluctuations of the $m$-th mode by $V_{\text{zpf},m}^{(M)} = \sqrt{2m+1}\sqrt{\hbar\omega^{(M)}_0/2C^{(M)}_0}$.\\
The interaction term $\hat{H}^{(M)}_{\text{int}}$ is given by
\begin{equation}
  \hat{H}^{(M)}_{\text{int}} = \sum_{m=0}^{m<M}\sum_{m'=m+1}^{m'<M}G^{(M)}_{m,m'}(a_m+a_m^\dagger)(a_{m'}+a_{m'}^\dagger)
  + \sum_{m=0}^{m<M}\hbar\bar{g}_{m}^{(M)}\hat{N}_J(a_m+a_m^\dagger)\ ,
\end{equation}
where $G_{m,m'}^{(M)}$ quantifies the coupling between the $m$-th and $m'$-th modes of the resonator through the presence of the capacitances introduced by the AA
\begin{equation}
  G_{m,m'}^{(M)} = -\frac{C_0C_cC_J}{MC_cC_J+C_0(C_c+C_J)} \frac{(C_0^{(M)})^2}{C_0^2}V_{\text{zpf},m}^{(M)}V_{\text{zpf},m'}^{(M)}\ ,
\end{equation}
and $\hbar\bar{g}_{m}^{(M)}= \beta^{(M)}V^{(M)}_{\text{zpf},m} 2e$ quantifies the coupling between the $m$-th mode of the resonator and the AA. It is weighted by the capacitance ratio
\begin{equation}
  \beta^{(M)} = \frac{C_0C_c}{MC_cC_J+C_0(C_c+C_J)} \frac{C_0^{(M)}}{C_0}\ .
\end{equation}
We can also write the Hamiltonian in the basis of eigenstates of the AA Hamiltonian. Defining the eigenstates $\{\ket{i}^{(M)}\}$ and eigenvalues $\epsilon_i^{(M)}$ by $\hat{H}^{(M)}_{\text{AA}}\ket{i}^{(M)} = \hbar\epsilon_i^{(M)}\ket{i}^{(M)}$ and making the transformation $\hat{H}^{(M)} \rightarrow \sum_i\ket{i}^{(M)}\bra{i}^{(M)}\hat{H}^{(M)}\sum_j\ket{j}^{(M)}\bra{j}^{(M)}$ we obtain the final form of the Hamiltonian

\begin{align}
\begin{split}
  \hat{H}^{(M)} &= \sum_{m=0}^{m<M}\hbar\omega^{(M)}_m\hat{a}_m^\dagger \hat{a}_m \\
  &+ \sum_i\hbar\epsilon_i^{(M)} \ket{i}^{(M)}\bra{i}^{(M)} \\
  &  \sum_{i,j}\sum_{m=0}^{m<M}\hbar g^{(M)}_{m,i,j}\ket{i}^{(M)}\bra{j}^{(M)}(\hat{a}_m+\hat{a}_m^\dagger)
   \\
  & \sum_{m=0}^{m<M}\sum_{m'=m+1}^{m<M}G^{(M)}_{m,m'}(\hat{a}_m+\hat{a}_m^\dagger)(a_{m'}+a_{m'}^\dagger)
  \end{split}
  \label{eq:circuit_Hamiltonian}
\end{align}
where the coupling $g_{m,i,j}$ is given by:
\begin{equation}
  \hbar g_{m,i,j} = V^{(M)}_{\text{zpf},m}2e\beta^{(M)}\bra{i}^{(M)}\hat{N}_J\ket{j}^{(M)}\ .
\end{equation}

If $C_J=0$, this Hamiltonian reduces to the one given in the main text. If not we can make use of a Bogoliubov transformation to express it in terms of the eigenmodes of the resonator as described in the Sec.~\ref{sec:Bogo}. This would allow us to recover the form of the Hamiltonian given in the main text. Alternatively the above hamiltonian can be diagonalized as it is to obtain an energy spectrum.

\begin{figure}[h!]
\centering
\includegraphics[width=0.22\textwidth]{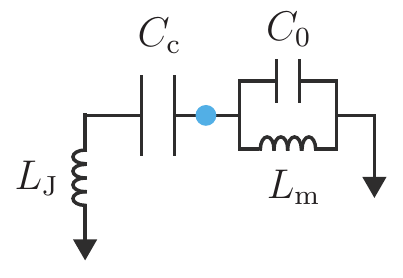}
\caption{Circuit of a linearized AA (series LC oscillator) coupled to a single mode $m$ of the resonator. Studying this circuit provides a good approximation of the Lamb shift in the Transmon regime $E_J\gg E_C$. The coupling point is shown as a blue dot.}
\label{fig:circuit_shift}
\end{figure}

\section{Dispersive shift of coupled LC oscillators}
In this section, we derive the Lamb shift of a linearized AA (\text{i.e.} a series LC oscillator) dispersively coupled to a single resonator mode for the case $C_J=0$ as shown in Fig. \ref{fig:circuit_shift}. The atom is linearized by discarding the purely non-linear part of the Josephson junction, leaving an inductor $L_J = \hbar^2/4e^2E_J$ \cite{Nigg2012}. We find that this shift gives a good approximation of the Lamb shift of high modes in the Transmon regime $E_J\gg E_C^{(0)}$. We denote by $L_J$ the inductance of the linearized AA and by $L_m$ and $C_0$ the inductance and capacitance of a coupled parrallel LC oscillator representing a bare resonator mode. The dispersive approximation assumes 
\begin{equation}
  \omega_m\gg\omega_a\ ,
  \label{eq:first_ass}
\end{equation} 
where $\omega_a$ is the resonance frequency of the bare linearized atom $\omega_a = 1/\sqrt{L_J C_c}$ and $\omega_m$ is the resonance frequency of the bare mode resonator $\omega_m = 1/\sqrt{L_m C}$. This condition is assumed to be met due to a small mode inductance
\begin{equation}
  L_J\gg L_m
  \label{eq:second_ass}
\end{equation} 
as is the case for high frequency modes $m\gg1$. Resonance is reached when the input impedance of the parallel LC oscillator is equal to minus that of the series LC oscillator, which is equivalent to a boundary condition of matching voltage and current at the coupling point shown in Fig. \ref{fig:circuit_shift}. This condition reads
\begin{equation}
  \frac{1}{iC_c\omega}+iL_J\omega = - \frac{1}{iC_0\omega+\frac{1}{iL_m\omega}}\ .
\end{equation}
Introducing the bare resonance frequencies corresponding to both resonators shunted to ground at the coupling point, this equation can be rewritten
\begin{equation}
  \omega^4-\omega^2\bigg(\omega_m^2+\omega_a^2 + \frac{L_m}{L_J}\omega_m^2\bigg)+\omega_m^2\omega_a^2=0\ .
\end{equation}
This equation has two positive solutions
\begin{equation}
  \omega_{\pm} = \frac{\omega_m}{\sqrt{2}}\sqrt{1+\eta\bigg(1+\frac{C_0}{C_c}\bigg)\pm\sqrt{\bigg(1+\eta\bigg(1+\frac{C_0}{C_c}\bigg)\bigg)^2-4\frac{C_0}{C_c}\eta}}\ ,
\end{equation}
where we introduced the quantity $\eta = L_m/L_J$. In the assumption of Eq. (\ref{eq:second_ass}), we obtain to first order in $\eta$ the resonance frequency
\begin{equation}
  \omega_-\simeq\omega_a-\frac{\omega_a}{2}\frac{L_m}{L_J}\ ,
  \label{eq:classical_lamb_shift_1}
\end{equation}
which yields the value of this shift $\bar{\chi}_m = -\frac{\omega_a}{2}\frac{L_m}{L_J}$. If we introduce the Josephson energy through $L_J = \hbar^2/4e^2E_J$, the atomic frequency $\hbar\omega_a = \sqrt{8E_JE_C^{(0)}}$ and the coupling $\hbar \gamma_m = 2e\sqrt{\frac{\hbar\omega_m}{2C_0}}\bigg(\frac{E_J}{32E_C^{(0)}}\bigg)^\frac{1}{4}$, this shift can be written in the language of the Rabi Hamiltonian following
\begin{equation}
  \bar{\chi}_m = -2\frac{\big(\omega_a\big)^2\big(\gamma_m\big)^2}{\omega_m^3}\ .
\end{equation}

Extrapolating this formula for the case of a non-linearized atom in the Transmon regime by making the approximatinos $\omega_a^{(0)}\simeq\omega_a$ and $g_m^{(0)} \simeq \gamma_m$ we obtain the formula for the shift presented in the main text
\begin{equation}
  \chi_m \simeq -2\frac{\big(\omega_a^{(0)}\big)^2\big(g_m^{(0)}\big)^2}{\omega_m^3}\ .
\end{equation}

\section{Numerical methods}
In order to perform numerical calculations, we first diagonalize the AA Hamiltonian $\hat{H}^{(M)}_{\text{AA}}$ (also known as Cooper pair box Hamiltonian) in the charge basis $\{\ket{N_J}\}_{N_J=-N_{\text{max}}, .., +N_{\text{max}}}$ where $\ket{N_J}$ is an eigenstate of $\hat{N}_J$. In this basis the Josephson junction term is given by (see Ref.~\cite{Schuster2007a})
\begin{equation}
  \cos (\hat{\delta}) = \frac{1}{2}\sum_{N=-\infty}^{+\infty} \ket{N_J}\bra{N_J+1} + \ket{N_J+1}\bra{N_J}\ .
\end{equation}
The basis is truncated to a certain number of Cooper pairs $\pm N_{\text{max}}$. We found that using more than $N_{\text{max}} = 20$ has little impact on simulation results for our set of example parameters. After diagonalization of $H_{\text{AA}}$ we can inject the values for $\epsilon_i^{(M)}$ and $\bra{i}^{(M)}\hat{N}_J\ket{j}^{(M)}$ into the Hamiltonian $\hat{H}^{(M)}$ which we in turn diagonalize. Numerical calculations are performed using the Python library QuTIP \cite{johansson2013qutip}.\\ 

\begin{figure}[h!]
\centering
\includegraphics[width=0.30\textwidth]{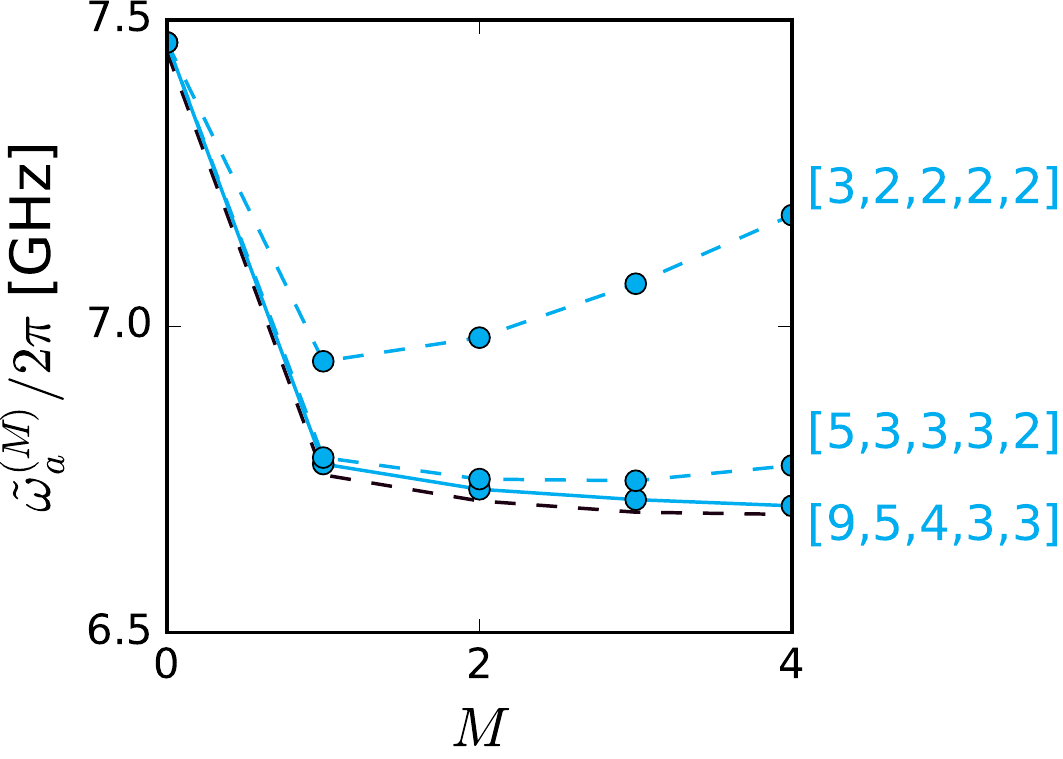}
\caption{The blue dots correspond to calculations of the dressed first AA transition frequency as we increase the number of modes included in the circuit model. Associated to each line is an array of integers, the first number of the array corresponds to the number of AA levels included in the model, and the following numbers correspond to the number of photon levels included, ordered with increasing mode frequency. Including less photon levels in the modes and in the AA leads to divergences. The dashed black line is the result of applying black box quantization to the $M$ mode lumped element equivalent circuit of the system.}
\label{fig:numerics}
\end{figure}

What must ensue is a careful choice of the size of the Hilbert space, namely the number of photon levels $n_m$ for the mode $m$ as well as the number of AA levels $n_a$. Note that the size of the Hilbert space scales as $2^{n_a}\prod_{m=0}^{m<M} n_{m}$. We find that a high number of photon levels are needed for convergence. This is particularly true for the modes which are the closest (in frequency) to $\omega^{(0)}_{ge}$. This is illustrated in Fig. \ref{fig:numerics} and explains the difficulty of providing a good estimate for the effective Lamb shift through a simple application of perturbation theory. \\

\section{Bogoliubov transformation} \label{sec:Bogo}

\begin{figure}[h!]
\centering
\includegraphics[width=0.35\textwidth]{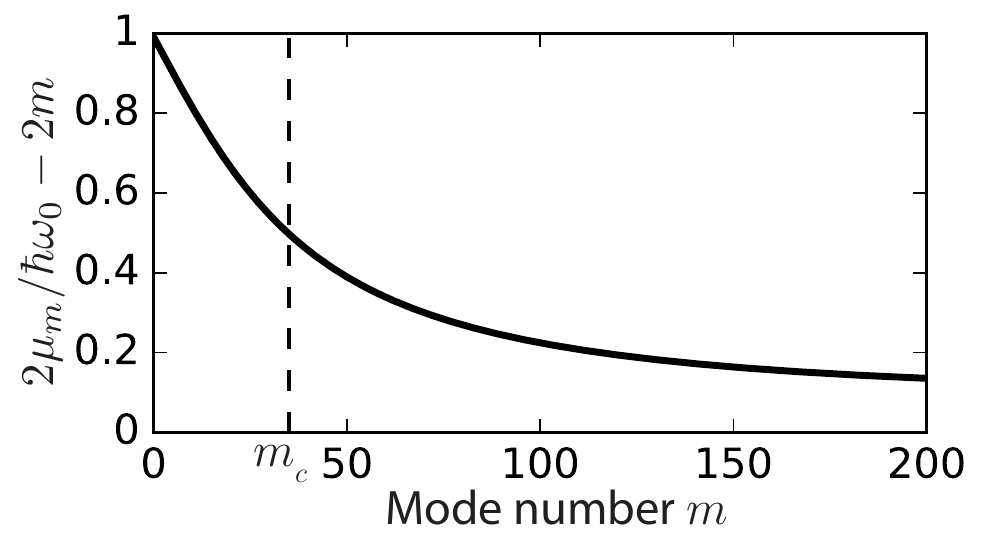}

\caption{Mode frequency as a function of the number of modes in the case $C_J=5$ fF (other circuit parameters fixed in the main text). The critical mode $m_C$ marks a transition from the regime where the resonator acts as a $\lambda/4$ resonator, with frequencies $2\mu_m\simeq (2m+1)\omega_0$ to a regime where the resonator becomes a $\lambda/2$ resonator, with eigenfrequencies $2\mu_m\rightarrow 2m\omega_0$.}
\label{fig:fn_vs_m}
\end{figure}

\begin{figure}[h!]
\centering
\includegraphics[width=0.2\textwidth]{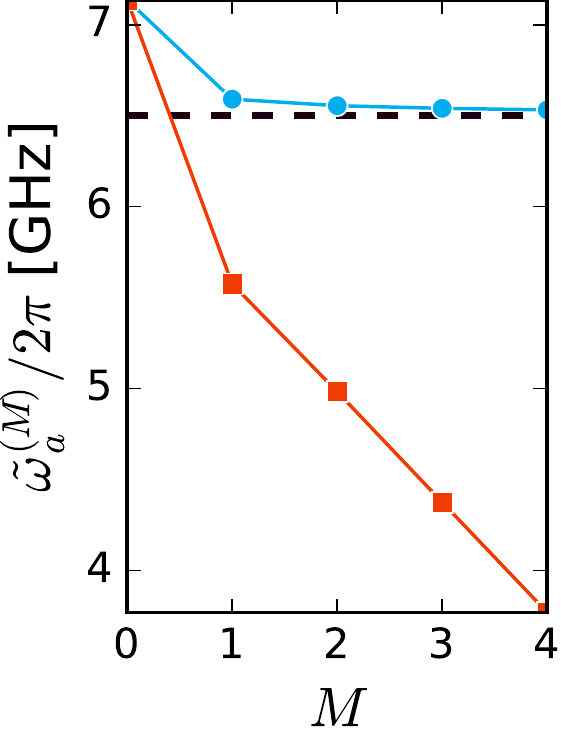}

\caption{Convergence of the spectrum in the case $C_J=5$ fF (other circuit parameters fixed in the main text). We plot the $\ket{g}\rightarrow\ket{e}$ transition frequency of the atom dispersively shifted by $M$ resonator modes as a function of $M$. Dots (squares) correspond to an exact diagonalization of the circuit (non-)renormalized multi-mode Rabi model. The frequency obtained by Black-box quantization (dashed line) provides a point of reference corresponding to the case when all modes are included.}
\label{fig:spectrum}
\end{figure}

In the case $C_J \ne 0$, one way to recover the form of the Hamiltonian presented in the main text is through a Bogoliubov transformation as described in this section. In the Hamiltonian given by Eq. (\ref{eq:circuit_Hamiltonian}), the energy of the bare resonator modes and the mode-mode coupling term correspond to the Hamiltonian of $M$ coupled harmonic oscillators
\begin{equation}
  \hat{H}'= \sum_{m=0}^{m<M}\hbar\omega^{(M)}_m\hat{a}_m^\dagger \hat{a}_m + \sum_{m=0}^{m<M}\sum_{m'=m+1}^{m<M}G^{(M)}_{m,m'}(\hat{a}_m+\hat{a}_m^\dagger)(a_{m'}+a_{m'}^\dagger)\ ,
  \label{eq:Hprime}
\end{equation}
which can be diagonalized through a Bogoliubov transformation even for $M$ on the order of thousands \cite{Javanainen1996}. We start by writting Eq. (\ref{eq:Hprime}) as follows
\begin{equation}
  \hat{H}'= \sum_{m,m'=0}^{m,m'<M}[\eta_{m,m'}(\hat{a}_m a_{m'} + \hat{a}_m^\dagger a_{m'}^\dagger) + \xi_{m,m'}(\hat{a}_m^\dagger a_{m'}+\hat{a}_m^\dagger a_{m'}^\dagger)]\ ,
\end{equation}
or, in matrix notation
\begin{equation}
  \hat{H}'= \boldsymbol{\alpha} ^T
\boldsymbol{h}'
\boldsymbol{\alpha}\ ,
  \end{equation}
where $\boldsymbol{\alpha}$ is a vector of the annihilation and creation operators
\begin{equation}
  \boldsymbol{\alpha}^T = [\hat{a}_0, \hat{a}_1, ..., \hat{a}_{M-1},\hat{a}_0^\dagger,\hat{a}_1^\dagger,...,\hat{a}_{M-1}^\dagger]\ ,
\end{equation}
and $\boldsymbol{h}'$ is the matrix
\begin{equation}
  \boldsymbol{h}'= \begin{bmatrix}
 \boldsymbol{\eta}     & \boldsymbol{\xi}     \\
 \boldsymbol{\xi}     &   \boldsymbol{\eta}     \\
\end{bmatrix}\ .
  \end{equation}
In this case $\eta_{m,m'} = \xi_{m,m'} = G_{m,m'}^{(M)}/2$ if $m \ne m'$ and $\eta_{m,m} = 0$, $ \xi_{m,m} = \hbar \omega_m^{(M)} /2$ otherwise. The challenge is now to find a matrix that maps $\alpha$ to a new set of creation and annihilation operators $\boldsymbol{\beta}$,
\begin{equation}
  \boldsymbol{\beta}^T = [\hat{b}_0, \hat{b}_1, ..., \hat{b}_{M-1},\hat{b}_0^\dagger,\hat{b}_1^\dagger,...,\hat{b}_{M-1}^\dagger] ,
\end{equation} 
which diagonalize $\hat{H}'$ whilst maintaining the expected commutation relations $[\hat{b}_m,\hat{b}_{m'}^\dagger] = \delta_{m,m'}$. Following the methodology described in Ref. \cite{Javanainen1996}, we introduce the matrix
\begin{equation}
  \boldsymbol{J}= 
\begin{bmatrix}
 \boldsymbol{0}     & \boldsymbol{I}     \\
 -\boldsymbol{I}     &   \boldsymbol{0}     \\
\end{bmatrix}\ ,
  \end{equation}
  where $\boldsymbol{I}$ ($\boldsymbol{0}$) is an $M\times M$ identity (zero) matrix. Diagonalizing the matrix $\boldsymbol{h}'\boldsymbol{J}$ yields eigenvalues that come in pairs such that if $\mu$ is an eigenvalue, then $-\mu$ is too. We order the eigenvalues and eigenstates such that the negative eigenvalues come first, in order of increasing absolute value, and the corresponding positive eigenvalues next, in the same order. We use the following notation for these eigenvalues
  \begin{equation}
    [-\mu_0,-\mu_1, ..., -\mu_{M-1}, \mu_0,\mu_1, ..., \mu_{M-1}]\ .
  \end{equation}
  We then construct a matrix $\boldsymbol{F}$ with the eigenvectors as columns and normalize them such that the $\boldsymbol{F}$ is sympletic: $\boldsymbol{F}^T\boldsymbol{JF}=\boldsymbol{J}$. To do so, we normalize each eigenvector $v_m$ such that $\sum_{i=0}^{i<2M}(v_m)_i^2 = 1$ and flip the sign of certain eigenvectors such that the first coeffecient of $v_m$ (with eigenvalue $-\mu_m$) has the same sign as the $M$-th coefficient of $v_{m+M}$ (with eigenvalue $\mu_m$). The matrix thus constructed should be of the form
\begin{equation}
  \boldsymbol{F}= 
\begin{bmatrix}
 \boldsymbol{A}     & \boldsymbol{B}     \\
 \boldsymbol{B}     &   \boldsymbol{A}     \\
\end{bmatrix}\ .
  \end{equation}
By defining the vector of annihilation and creation operators $\boldsymbol{\beta}$ as 
\begin{equation}
  \boldsymbol{\alpha} = \begin{bmatrix}
 \boldsymbol{A}     & -\boldsymbol{B}     \\
- \boldsymbol{B}     &   \boldsymbol{A}     \\
\end{bmatrix}\boldsymbol{\beta}\ , 
\end{equation}
we have defined a basis which diagonalizes $\hat{H}'$
 \begin{equation}
  \hat{H}'= \sum_{m=0}^{m<M}2\mu ^{(M)}_m\hat{b}_m^\dagger \hat{b}_m\ ,
\end{equation}
the new eigenenergies in fact being given by twice the positive eigenvalues of the previously diagonalized matrix. In this basis, the atom-mode interaction term becomes
\begin{equation}
  \hat{H}^{(M)}_{\text{int}} = \sum_{i,j}\sum_{m=0}^{m<M}\bigg[\sum_{m'=0}^{m'<M}\hbar g^{(M)}_{m',i,j}(\boldsymbol{A}-\boldsymbol{B})_{m',m}\bigg]\ket{i}^{(M)}\bra{j}^{(M)}(\hat{b}_m+\hat{b}_m^\dagger)\ ,
\end{equation}
and we recover the extended Rabi Hamiltonian structure by defining the coupling as
\begin{equation}
  g^{(M)}_{m,i,j} = \sum_{m'=0}^{m'<M}\hbar g^{(M)}_{m',i,j}(\boldsymbol{A}-\boldsymbol{B})_{m',m}\ .
\end{equation}
This coupling strength was plotted in the main text. In Fig.~\ref{fig:fn_vs_m}, we plot the frequencies of the newly defined eigenmodes of the resonator. As expected, these transition from the eigenfrequencies of a $\lambda/4$ resonator to those of a $\lambda/2$. In Fig. \ref{fig:spectrum}, we show the same plot as in Fig. 3 of the main text but for $C_J=5$ fF, the result of diagonalizing the Hamiltonian derived above.

\section{Dressing of the atomic charging energy}

  The Hamiltonian $\hat{H}^{(M+1)}$ with $M+1$ bosonic modes coupled to the Josephson junction (Eq. (1) of the Letter) lives in the Hilbert space $\hat{H}^{(M+1)}=\hat{H}_{N_J}\otimes_{m \leq M}\hat{H}_m$
  \begin{eqnarray}
  \hat{H}^{(M+1)}&=& \hbar\omega_M a_M^{\dagger}a_M+\hbar\bar{g}_M(a_M+a_M^{\dagger})\hat{N}_J + \hat{H}^{(M)}\\
  &=&\hbar\omega_M \left(a_M^{\dagger}+\frac{\bar{g}_M}{\omega_M}\hat{N}_J\right)\left(a_M+\frac{\bar{g}_M}{\omega_M}\hat{N}_J\right) + \hat{H}^{(M)}-\hbar\frac{\bar{g}_M^2}{\omega_M}\hat{N}_J^2\\
  &=& \hbar\omega_M b_M^\dagger b_M + \hat{\textbf{H}}^{(M)},
  \end{eqnarray}
  where we have defined the bosonic operators 
  \begin{eqnarray}
  b_M &=& a_M + \frac{\bar{g}_M}{\omega_M}\hat{N}_J,\\
  b_M^\dagger &=& a_M^\dagger + \frac{\bar{g}_M}{\omega_M}\hat{N}_J,
  \end{eqnarray}
  and the Hamiltonian
\begin{equation}
  \hat{\textbf{H}}^{(M)} = \hat{H}^{(M)}-\hbar\frac{\bar{g}_M^2}{\omega_M}\hat{N}_J^2
\end{equation}
  We look for an effective Hamiltonian which approximates the low energy part of $\hat{H}^{(M+1)}$. The pair $b_M$, $b_M^\dagger$ are canonically conjugate, $\left[b_M,b_M^\dagger\right]=1$. Thus, $b_M^\dagger b_M$ is a number operator. If $\hbar\omega_M$ is much larger than the characteristic energy of the low energy sector of $\hat{\textbf{H}}^{(M)}$, the low energy sector of $\hat{H}^{(M+1)}$ will be well approximated by setting $b_M^\dagger b_M$ to zero. That is, by studying the restriction of $\hat{H}^{(M+1)}$ to the vacuum subspace of $b_M$, namely 
  \begin{eqnarray}
  \mathcal{S}^{(M+1)}=\{\Ket{\Psi}/ \,\,\,b_M\Ket{\Psi}=0\}.
  \end{eqnarray}
  In order that the separation of scales that has been assumed indeed holds, it is also imperative that $\hat{\textbf{H}}^{(M)}$ acting on $\mathcal{S}^{(M+1)}$ results in states neighbouring $\mathcal{S}^{(M+1)}$. That is to say, that the commutator $\left[b_M,\hat{\textbf{H}}^{(M)}\right]$ acting on $\mathcal{S}^{(M+1)}$ be small. In the case at hand, 
  
  \begin{eqnarray}
    \left[b_M,\hat{\textbf{H}}^{(M)}\right]&=&\frac{\bar{g}_M}{\omega_M}\left[\hat{N}_J,\hat{H}^{(M)}\right]\\
    &=&i\frac{E_J\bar{g}_M}{\omega_M}\sin(\varphi_J),
  \end{eqnarray}
  so if $E_J, \bar{g}_M\ll \omega_M$ then we can say that the commutator above is small, and that $\hat{\textbf{H}}^{(M)}|_{\mathcal{S}^{(M+1)}}$ will provide a good effective Hamiltonian for $\hat{H}^{(M+1)}$. Notice that these conditions are increasingly better fulfilled with growing mode number $M$ for the model in the main text. We now construct explicitly the effective Hamiltonian $\hat{\textbf{H}}^{(M)}|_{\mathcal{S}^{(M+1)}}$. The subspace $\mathcal{S}^{(M+1)}$ can be expanded in the following basis
  \begin{equation}
  \Ket{\alpha_{N_J}}^{(M+1)} = \Ket{N_J}\Ket{\beta}^{(M)}\Ket{z_M = -\gamma_M N_J},
  \end{equation}
  where vectors $\Ket{\beta}^{(M)}$ form a basis of the truncated subspace $\otimes_{m< M}\hat{H}_m$, $\Ket{z_{M}}$ is a coherent state for the $(M+1)$-th mode and $\gamma_M = \bar{g}_M/\omega_M$. The original bosonic $a_M$ and Cooper-Pair number $\hat{N}_J$ operators act on this basis as
  \begin{eqnarray}
  a_M\Ket{\alpha_{N_J}}^{(M+1)}&=&-\gamma_M N_J\Ket{\alpha_{N_J}}^{(M+1)},\\
  \hat{N}_J\Ket{\alpha_{N_J}}^{(M+1)}&=&N_J\Ket{\alpha_{N_J}}^{(M+1)}.
  \end{eqnarray}
  Thus, the matrix elements of $\hat{H}^{(M+1)}|_{\mathcal{S}^{(M+1)}}$ are
  \begin{eqnarray}
  \Bra{\alpha_{N_J}}^{(M+1)}\hat{H}^{(M+1)}\ket{\alpha_{M_J}}^{(M+1)}&=& \left<-\gamma_N N_J|-\gamma_N M_J\right>\bra{\alpha_{N_J}}^{(M)}\hat{\textbf{H}}^{(M)}\ket{\alpha_{M_J}}^{(M)}\\
  &=&e^{-\gamma_M^2 \left(N_J-M_J\right)^2/2}\bra{\alpha_{N_J}}^{(M)}\hat{\textbf{H}}^{(M)}\ket{\alpha_{M_J}}^{(M)}\\
  &\approx&\bra{\alpha_{N_J}}^{(M)}\hat{\textbf{H}}^{(M)}\ket{\alpha_{M_J}}^{(M)},
  \end{eqnarray}
  where the last line gives us a further approximation, valid if $\gamma_M\ll1$ and the low energy states of $\hat{\textbf{H}}^{(M)}$ have small dispersion for $\hat{N}_J$. If these indeed hold, $\hat{\textbf{H}}^{(M)}$ itself is a good effective Hamiltonian for $\hat{H}^{(M+1)}$. We can iterate this procedure down to a mode $L$ for which the above conditions still holds. For $M\rightarrow\infty$, an effective Hamiltonian with $L$ modes is then given by 

  \begin{equation}
  \hat{H}=\sum_{m=0}^{m<L}\hbar\omega_{m}^{(L)}a_{m}^{\dagger}a_{m}+ \hbar\bar{g}_{m}^{(L)}\left(a_{m}+a_{m}^{\dagger}\right)\hat{N}_{J}+4\tilde{E}_{C}^{(L)}\hat{N}_{J}^{2}-E_{J}\,\mathrm{cos}(\hat{\delta}),
  \end{equation}
\begin{equation}
\tilde{E}_C^{(L)} = \lim_{M\rightarrow\infty}E_C^{(M)} - \sum_{m\geq L}^{M}\hbar(\bar{g}^{(M)}_{m})^{2}/4\omega_{m} .
\end{equation}

\bibliography{library_SI,extra_SI}